# New record of high *ZT* found in hybrid transition-metal-dichalcogenides

Yulou Ouyang[1], Yuee Xie[1], Zhongwei Zhang[1], Qing Peng[2*], and Yuanping Chen[1]*

[1] Department of Physics, Xiangtan University, Xiangtan 411105, Hunan, P.R. China

[2] Department of Mechanical, Aerospace and Nuclear Engineering, Rensselaer Polytechnic Institute, Troy, NY, 12180, USA

**ABSTRACT:** The search for thermoelectrics with higher figures of merit (*ZT*) will never stop due to the demand of heat harvesting. Single layer transition metal dichalcogenides (TMD), namely $MX_2$ (where M is a transition metal and X is a chalcogen) that have electronic band gaps are among the new materials that have been the focus of such research. Here, we investigate thermoelectric transport properties of hybrid armchair-edged TMDs nanoribbons, by using the nonequilibrium Green's function technique combined with the first principles and molecular dynamics methods. We find a *ZT* as high as 7.4 in hybrid $MoS_2$/$MoSe_2$ nanoribbons at 800K, creating a new record for *ZT*. Moreover, the hybrid interfaces by substituting X atoms are more efficient than those by substituting M atoms to tune the *ZT*. The origin of such a high *ZT* of hybrid nanoribbons is the high density of the grain boundaries: the hybrid interfaces decrease thermal conductance drastically without a large penalty to electronic conductance.

Corresponding authors: chenyp@xtu.edu.cn, qpeng.org@gmail.com.

# I. INTRODUCTION

With the success of graphene, monolayer transition metal dichalcogenides (TMDs) in the form of $MX_2$ (M is a transition metal while X is a chalcogen), have attracted much attention in various fields[1-3] because of their excellent electronic and optoelectronic properties.[4-7] For example, the carrier mobility of a $MoS_2$ transistor can approach $200 cm^2V^{-1}S^{-1}$, and its on/off ratio approaches $10^8$ at room temperature.[8] High-performance light-emitting diodes and field-effect transistors based on $WS_2$ have also been reported.[9] Contrasting to the electronic properties, the thermal properties of TMDs are still not well known. Recent studies showed that the thermal conductivities of TMDs may be much lower than that of graphene.[10,11] In fact, theoretical calculations indicated that the thermal conductivity of $MoS_2$ is smaller than that of graphene by three orders,[12] while the former reported by experiment is also only 0.01~0.5 of the latter.[13] Therefore, TMDs possess high electron conductivity and low thermal conductivity simultaneously.

Thermoelectric materials converting heat to electricity have been a long term focus.[14] It is well known that the efficiencies of thermoelectric materials are measured by the so-called figure of merit *ZT*, defined as $ZT = S^2\sigma T/k$, where $\sigma$ is electronic conductance, $k$ is total thermal conductance, and $S$ is Seebeck coefficient.[15,16] A good thermoelectric material should have high electron conductivity and low thermal conductivity. As small band gap semiconductors, TMDs are good candidates for thermoelectric materials that have high *ZT* values. Hippalgaonkar *et al.*[17] and Wu *et al.*[18] experimentally showed that single and few-layer $MoS_2$ have high thermoelectric power factor. A previous calculation based on first-principles and Boltzmann transport theory showed that the *ZT* value of single layer $MoS_2$[19] can be 0.11 at 500K and $MoS_2$ armchair nanoribbons[20] can be optimized to 3.5 at room temperature. A first-principles calculation on single-layer or few-layer TMD indicated that bilayer MoSe$_2$ gives a maximum *ZT* value of 2.4.[21]

Interestingly, with the development of nanotechnology, vertical and in-plane hybrid heterostructures from TMD monolayers have been fabricated.[22,23] For the in-plane hybrid heterostructures, there exist hybrid interfaces of two TMD structures, such as MoS$_2$/WSe$_2$ and MoS$_2$/WS$_2$.[24,25] The interfaces would have a great effect on the thermal transport because the scattering on interfaces reduces the mean free path of phonons.[26,27] On the contrary, the effect of interfaces on the electron transport is relatively smaller, because TMDs have similar electronic properties. Therefore, one hypothesis is that the hybrid TMD structures may have higher *ZT* than the pristine ones, which is one of the extended goals of this study.

In this paper, we investigate thermoelectric properties of hybrid TMD nanoribbons using a ballistic transport approach. The hybrid armchair-edged nanoribbons show high *ZT* due to the fact that the hybrid interfaces reduce thermal conductance drastically without a large penalty to electronic conductance. The *ZT* values of pristine armchair TMD nanoribbons are approximately 1.5 ~ 2.1 at room temperature, while those of hybrid nanoribbons are 2 ~ 3 times that of pristine ones, depending on the number of hybrid interfaces. For example, the *ZT* of the hybrid nanoribbons MoS$_2$/MoSe$_2$ or MoS$_2$/WS$_2$ with three interfaces is around 3.0 while that of pristine MoS$_2$ is only 1.5. With the increase of temperature and interface number, the *ZT* values are considerably improved. Our calculations indicate that the highest *ZT* of hybrid nanoribbons can approach 7.4 at a temperature of 800 K. Moreover, the value is expected to be further increased by doping, edge defects, and adsorption. Therefore, the hybrid TMD nanoribbons may have promising applications in thermal energy harvesting.

## II. MODEL AND METHOD

Figure 1 shows the hybrid armchair-edged TMD nanoribbons $M_1(X_1)_2/M_2(X_2)_2$ we studied ($M_1$ and $M_2$ are transition metals while $X_1$ and $X_2$ are chalcogens), where the $M_1(X_1)_2$ and $M_2(X_2)_2$ form a finite superlattice and the periodic number is labelled as $N$. As a transport system, the finite superlattice is divided into three regions, one central scattering region and two leads. The length of the central region is labelled as $L$ while that of a period is labelled as $l$, and thus $L = N \times l$. The widths of the nanoribbons are labelled as $W$. Here, we consider three typical hybrid nanoribbons $MoS_2/MoSe_2$, $MoS_2/WS_2$, and $MoS_2/WSe_2$. The former two are hybrid structures formed by substituting X and M atoms, respectively, while the last one is formed by substituting both X and M atoms. It is noted that, in $MX_2$, the geometrical differences between the structures with the same X are smaller than those with the same M. For example, the lattice constants of $MoS_2$ and $WS_2$ are 3.11 and 3.13 Å, while $MoSe_2$ and $WSe_2$ are 3.24 and 3.25 Å, respectively.[4]

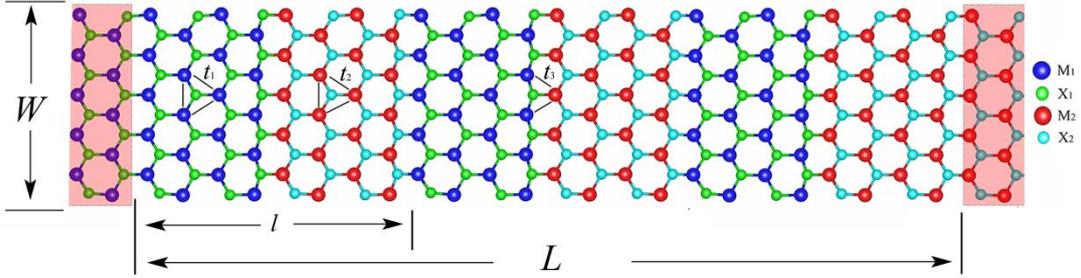

**Figure 1| Hybrid TMD armchair-edged nanoribbons $M_1(X_1)_2/M_2(X_2)_2$.** Left and right (red) parts are the two leads, while the middle part is a central scattering region which is a finite superlattice. The width and length of the central scattering region is labelled as $L$ and $W$, respectively. The periodic length of the finite superlattice is labelled as $l$, and thus $L = N \times l$ ($N$ represents the periodic number, and here $N = 3$). $t_1$, $t_2$, and $t_3$ represent the hopping integrals between $M_1$-$M_1$, $M_2$-$M_2$, and $M_1$-$M_2$ atoms, respectively. The structural parameters are $l = 2.21$ nm and $W = 1.53$ nm.

..................................................................................................................................................

We use the non-equilibrium Green's function (NEGF) method to calculate the thermoelectric transport properties of these hybrid nanoribbons.[28-30] At first, we consider their electronic properties. To describe the electronic properties, a tight-binding (TB) Hamiltonian is used, [31,32]

$$H = H_1 + H_2 + H_{12}, \qquad (1)$$

where $H_i = \sum_l \varepsilon_i a_l^\dagger a_l + \sum_{lm} t_i a_l^\dagger a_m$ represents the Hamiltonians of the nanoribbons $M_i(X_i)_2$, while $H_{12} = \sum_{lm} t_3 a_l^\dagger b_m$ represents the interaction of $M_1(X_1)_2$ and $M_2(X_2)_2$. $t_i$ and $\varepsilon_i$ ($i = 1,2$) are the third-order hopping integrals for the nearest-neighbor atoms and site energies in $M_i(X_i)_2$, which can be obtained from the GGA parameters in Ref.24, while the hopping integral $t_3$ for the interaction of atoms $M_1$ and $M_2$ are taken to be the average values of $t_1$ and $t_2$ ($t_3 = \sqrt{t_1 \times t_2}$).[33-36] Based on the Hamiltonian in Eq. (1), the retarded Green's function is expressed as[37]

$$G^r(E) = [(E + i0^+)I - H_C - \Sigma_L^r - \Sigma_R^r]^{-1}, \qquad (2)$$

where $E$ is the electron energy, $I$ is the identity matrix, $H_C$ is the Hamiltonian of the central scattering region, $\Sigma_\beta^r = H^{C\beta} g_\beta^r H^{\beta C}$ denotes the self-energy of the left ($L$) and right ($R$) regions, and $g_\beta^r$ is the surface Green's function of the leads which can be computed by a recursive iteration technique.[38] Once the retarded Green's function is obtained, we can calculate the electronic transmission coefficient

$$T_e(E) = Tr[G^r(E)\Gamma_L G^a(E)\Gamma_R], \qquad (3)$$

where $G^a(E) = [G^r(E)]^+$ is the advanced Green's function, and $\Gamma_\beta = i(\Sigma_\beta^r - \Sigma_\beta^a)$ describes the interaction between the left or right lead and the central region. Once $T_e(E)$ is obtained, the electronic conductance $\sigma$, Seebeck coefficient $S$, and electronic thermal conductance $k_e$ can be calculated,[36,39]

$$\sigma(\mu,T) = e^2 L_0(\mu,T), \tag{4}$$

$$S(\mu,T) = \frac{1}{eT} \frac{L_1(\mu,T)}{L_0(\mu,T)}, \tag{5}$$

$$k_e(\mu,T) = \frac{1}{T}\left[L_2(\mu,T) - \frac{L_1(\mu,T)^2}{L_0(\mu,T)}\right], \tag{6}$$

where $L_n(\mu,T) = \frac{2}{h}\int_{-\infty}^{+\infty}(E-\mu)^n\left[-\frac{\partial f_e(E,\mu,T)}{\partial T}\right]T_e(E)dE$ is the Lorenz integral and $f_e(E,\mu,T)$ is the Fermi-Dirac distribution function at the chemical potential $\mu$ and temperature $T$.

The phonon transmission coefficient $T_p(\omega)$ can also be calculated by the NEGF method analogous to that of calculating electronic transmission coefficients. In Eq. (2) one only needs to change $E$ into $\omega^2$, $H_C$ into $K_C$, and correspondingly compute the self-energy, and then get the phonon retarded Green's function, where $\omega$ is the vibrational frequency of phonons and $K_C$ is the force constant matrix of the central region, while $K_C$ can be obtained by the dynamics software "General Utility Lattice Program" (GULP)[40] according to the Stillinger-Weber (SW) potential functions.[41] Here, the Stillinger-Weber (SW) potential parameters are used to describe the inter-atomic interactions in the hybrid nanoribbons,[42]

$$\Phi(1,\dots,N) = \sum_{i<j} V_2(i,j) + \sum_{i<j<k} V_3(i,j,k), \tag{7}$$

$$V_2 = \epsilon A\left(B\sigma^p r_{ij}^{-p} - \sigma^q r_{ij}^{-q}\right)e^{[\sigma(r_{ij}-a\sigma)^{-1}]}, \tag{8}$$

$$V_3 = \epsilon \lambda e^{\left[\gamma\sigma(r_{ij}-a\sigma)^{-1}+\gamma\sigma(r_{jk}-a\sigma)^{-1}\right]}\left(\cos\theta_{jik} - \cos\theta_0\right)^2, \tag{9}$$

where $V_2$ and $V_3$ represent two-body and three-body interactions, and $\theta_0$ is the initial angle. All the parameters in Equation (6) and (7) are obtained by fitting the phonon dispersions calculated from first-principles methods, of which the detailed processes are presented in the Supplementary Material (SII). Then, in GULP, the force constant $Kc$ can be given by the second derivatives with respect to the potential energy. As $T_p(\omega)$ is calculated, the phonon thermal conductance can be given by $k_p(T) = \frac{h}{4\pi^2}\int_0^\infty \frac{\partial f_p(\omega)}{\partial T} T_p(\omega)\omega d\omega$,[43] where $f_p(\omega)$ is the Bose-Einstein distribution function. Finally, the total thermal conductance $k = k_p + k_e$ is obtained. In general, the best thermoelectric performance of one structure does not occur at the intrinsic chemical potential, i.e., $\mu = E_f$ ($E_f$ is the Fermi level). To explore its thermoelectric properties, one need to investigate its $ZT$ values at different $\mu$.[44-46]

## III. RESULTS AND DISCUSSION

We firstly show $ZT$ as a function of chemical potential $\mu$ at $T$ = 300 K for four armchair-edged pristine nanoribbons $MoSe_2$, $WSe_2$, $MoS_2$, and $WS_2$, in Figure 2(a). The maximum values of $ZT$ (Max $ZT$) for the four nanoribbons all appear at $\mu < 0$. The Max $ZT$ of $MoSe_2$ (1.8) and $WSe_2$ (2.1) are larger than those of $MoS_2$ (1.5) and $WS_2$ (1.5). The $ZT$ values for the three hybrid nanoribbons $MoS_2/MoSe_2$, $MoS_2/WS_2$, and $MoS_2/WSe_2$ are shown in Figure 2(b) and 2(c). Figure 2(b) corresponds to the simplest hybrid nanoribbons whose central scattering region includes one period ($N$ = 1), i.e., there is only one hybrid interface in the structure, while Figure 2(c) corresponds to the hybrid nanoribbons with $N$ = 2 in which there are three interface

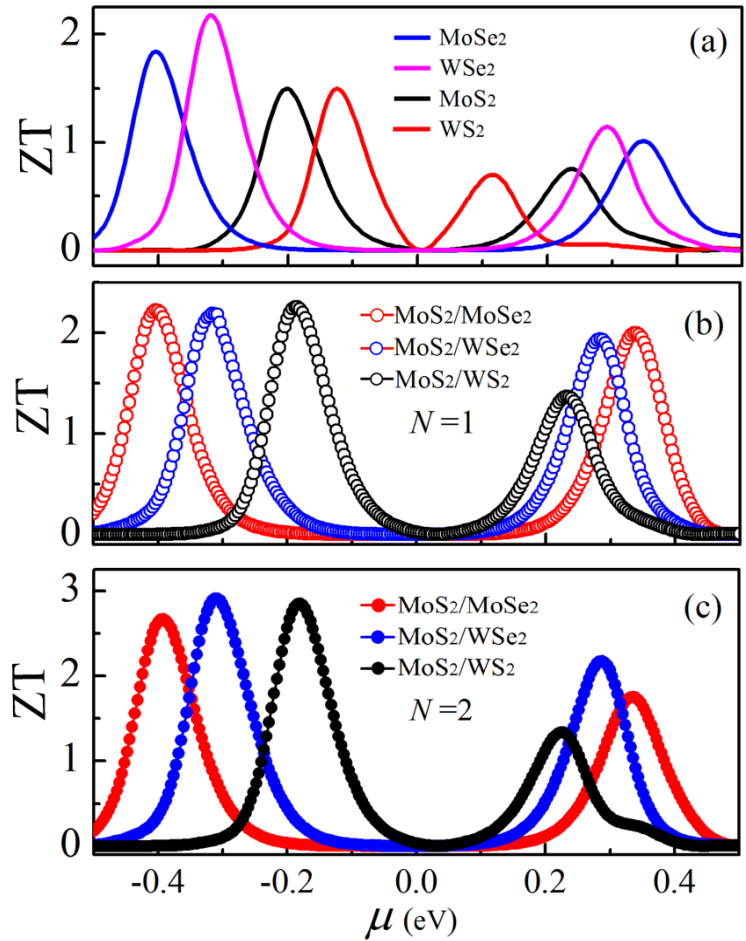

**Figure 2| Chemical potential and interface density effect** (a) *ZT* as a function of chemical potential $\mu$ for pristine nanoribbons $MoSe_2$, $WSe_2$, $MoS_2$, and $WS_2$, and respectively, at *T* = 300 K. (b) *ZT* as a function of chemical potential $\mu$ for hybrid nanoribbons $MoS_2/MoSe_2$, $MoS_2/WSe_2$, and $MoS_2/WS_2$ with one interface (*N* = 1). (c) *ZT* as a function of chemical potential $\mu$ for hybrid nanoribbons $MoS_2/MoSe_2$, $MoS_2/WSe_2$, and $MoS_2/WS_2$ with three interfaces (*N* = 2).

…………………………………………………………………………………………..

in their central scattering region (Note that the number of interfaces is equal to 2*N*-1). It is found that the Max *ZT* of all hybrid nanoribbons are higher than those of pristine nanoribbons. The hybrid nanoribbons with one interface have a Max *ZT* of approximately 2.1, while those with three interfaces have a Max *ZT* greater than 2.7. This indicates that the hybrid interfaces enhance thermoelectric properties. Moreover, the more the interfaces have, the higher the *ZT* values are.

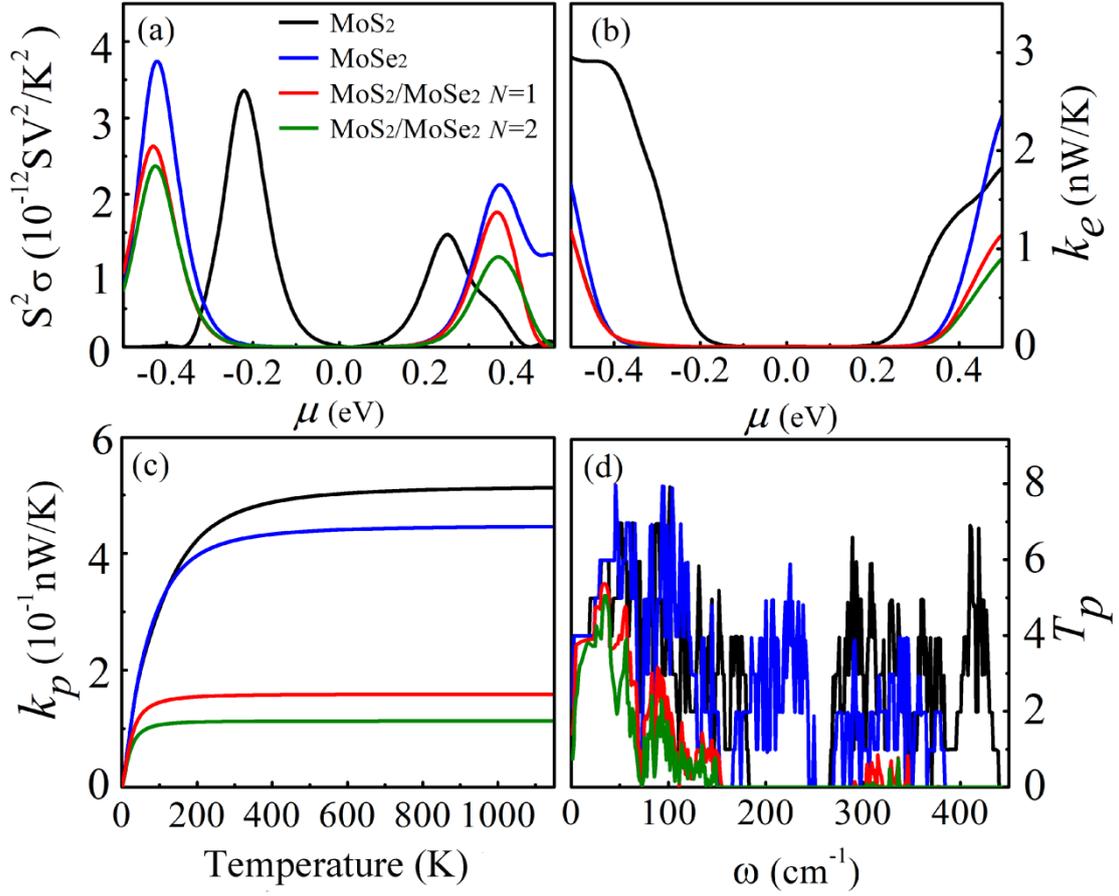

**Figure 3|** (a) $S^2\sigma$, (b) $k_e$ of $MoS_2$, $MoSe_2$, $MoS_2/MoSe_2$ ($N$ = 1), and $MoS_2/MoSe_2$ ($N$ = 2) nanoribbons as a function of chemical potential $\mu$, respectively, at $T$ = 300 K. (c) $k_p$ of $MoS_2$, $MoSe_2$, $MoS_2/MoSe_2$ ($N$ = 1), and $MoS_2/MoSe_2$ ($N$ = 2) nanoribbons as a function of temperature $T$. (d) Phonon transmission coefficients $T_p$ of $MoS_2$, $MoSe_2$, $MoS_2/MoSe_2$ ($N$ = 1), and $MoS_2/MoSe_2$ ($N$ = 2) nanoribbons as a function of frequency.

……………………………………………………………………………...

To explore the effect of hybrid interfaces on the thermoelectric properties, we compare the electronic and thermal properties of pristine and hybrid nanoribbons in Figure 3. Thermal power $S^2\sigma$ and thermal conductance $k_e$ and $k_p$ are the three factors to determine $ZT$ value. At the room temperature, $S^2\sigma$ and $k_e$ are only functions of chemical potential $\mu$, while $k_p$ is a constant. Figures 3(a-b) compare $S^2\sigma$ and $k_e$ of the nanoribbons $MoS_2$, $MoSe_2$, and $MoS_2/MoSe_2$ with $N$ = 1 and 2, while Figure 3(c) compares $k_p$ of these nanoribbons as a function of temperature $T$. Seen from Figures 3(a-b), one can find that the curves of the MoSe$_2$ nanoribbons are different from those of the MoS$_2$ nanoribbons, which originate from their different band gaps. Because the band gap of the MoSe$_2$ nanoribbons is larger than that of MoS$_2$ nanoribbons, the threshold values of $S^2\sigma$ for the former are larger than the latter. It is noted that the Max $ZT$ of MoS$_2$ and MoSe$_2$ appear at $\mu$ = -0.2 and -0.4 eV (see Figure 2(a)), respectively. In addition, the $S^2\sigma$, $k_e$, and $k_p$ of MoS$_2$ at $\mu$ = -0.2 eV have little difference with those of MoSe$_2$ at $\mu$ = -0.4 eV. Relatively, $S^2\sigma$ of the former is smaller while $k_p$ is larger, and thus the Max $ZT$ of the MoSe$_2$ nanoribbons is higher than that of MoS$_2$ nanoribbons. For the hybrid structures, their energy gaps should be equal to the wider one, i.e., equal to that of MoSe$_2$ nanoribbons. Therefore, the $S^2\sigma$ and $k_e$ of the hybrid nanoribbons have the same thresholds as that of the MoSe$_2$. Figures 3(a-c) illustrate that, although both electron and thermal transport are weakened by the scattering at hybrid interfaces, the effect of interfaces on the thermal transport is much larger than that on the electron transport. The maximums of $S^2\sigma$ are decreased by less than 1.5 times by the interfaces while the $k_p$ are decreased by approximately 2.5 times. The $T_p$ spectra in Figure 3(d) clearly show that the interfaces reduce the phonon transmission. From these results, one can conclude that the improved thermoelectric properties of the hybrid structures originate mainly from the sharp decrease of phonon thermal conductance $k_p$.

It is known that $ZT$ is a function of temperature $T$. It is critical to know the working temperature that has the highest $ZT$. The Max $ZT$ of the three hybrid nanoribbons MoS$_2$/WS$_2$, MoS$_2$/WSe$_2$, and MoS$_2$/MoSe$_2$ with the variation of $T$ are shown in Figure 4(a). One can find that the Max $ZT$ increases at low temperature to

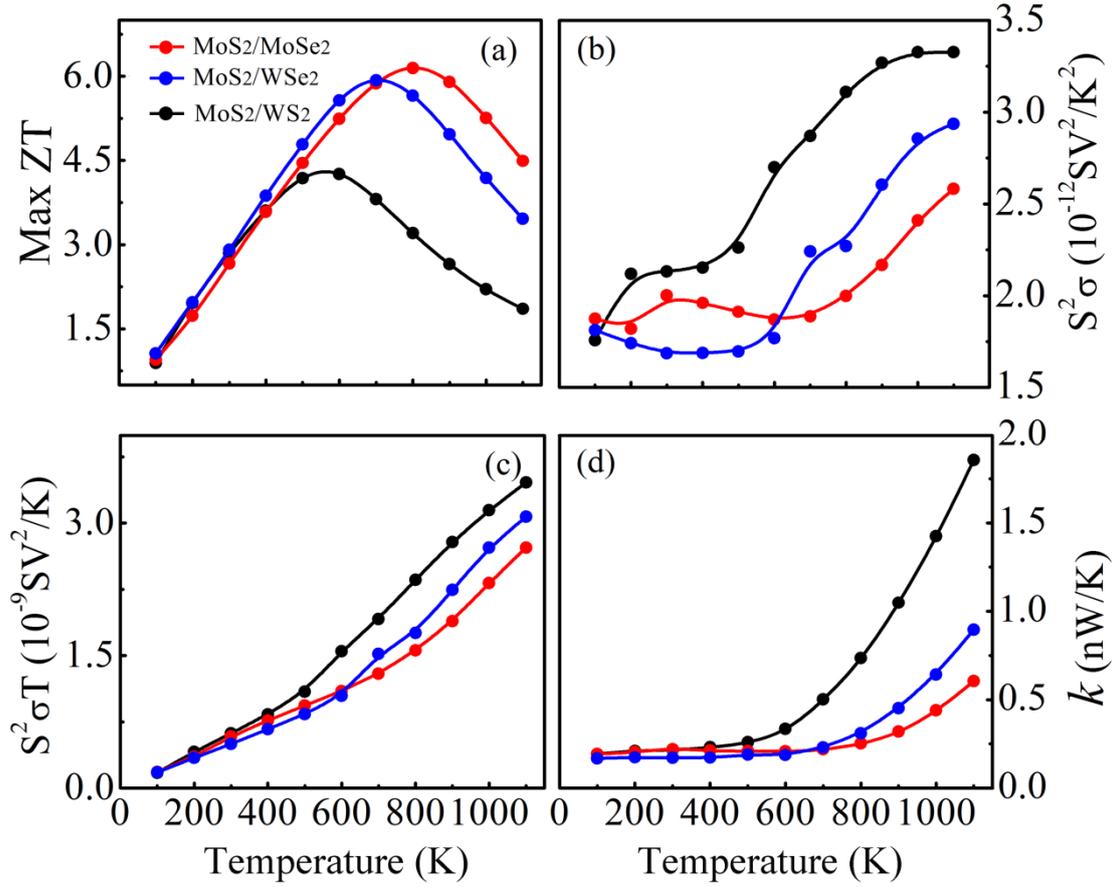

**Figure 4|** (a) The Max *ZT* values of hybrid nanoribbons MoS$_2$/MoSe$_2$, MoS$_2$/WSe$_2$, and MoS$_2$/WS$_2$ with $N = 2$ as functions of temperature *T*. (b) $S^2\sigma$, (c) $S^2\sigma T$, and (d) $k$, as a function of temperature *T* for the three hybrid nanoribbons corresponding to Max *ZT* in (a).

………………………………………………………………………………………..

the maximum followed by a reduction at high temperature. When *T* < 450K, all hybrid structures nearly have the same Max *ZT* values at the same temperature. When *T* > 450K, the difference in Max *ZT* of the three structures becomes apparent with the temperature. The highest *ZT* of MoS$_2$/WS$_2$, MoS$_2$/WSe$_2$, and MoS$_2$/MoSe$_2$ appear at *T* = 600, 700, and 800 K, respectively. Moreover, one can find that the Max *ZT* of MoS$_2$/MoSe$_2$ can approach 6.1 while that of MoS$_2$/WS$_2$ is 4.2. Although both of them represent very high energy conversion efficiencies, the hybrid interfaces formed by substituting X atoms are more efficient than those formed by substituting M atoms.

To explain the effect of temperature on the thermoelectric transport in the hybrid nanoribbons, we calculate the variations of $S^2\sigma$, $S^2\sigma T$, and $k$ corresponding to Figure 4(a), as shown in Figures 4(b-d). All thermal powers $S^2\sigma$ of the three hybrid nanoribbons increase approximately with the temperature (see Figure 4(b)), while the relation of all $S^2\sigma T$ and $T$ is nearly linear, and the difference of the three structures is small (see Figure 4(c)). Therefore, the transitions of Max $ZT$ in Figure 4(a) are induced by the variation of thermal conductance $k$, as shown in Figure 4(d). When $T$ < 450 K, each $k$ of the three hybrid structures are nearly the same constant, which is the reason why Max $ZT$ increases at low temperature and all hybrid structures nearly have the same Max $ZT$ values at the same temperature. When $T$ > 450K, $k$ of the three structures increases sharply, especially $MoS_2/WS_2$. For $MoS_2/WS_2$, $MoS_2/WSe_2$, and $MoS_2/MoSe_2$, the transitions of $k$ appear at $T_c$ = 600, 700, and 800 K, respectively. As a result, there are highest $ZT$ appearing at the $T_c$.

The thermal conductance $k$ of different hybrid nanoribbons are much different at high temperatures, as illustrated in Figure 4d. Because $k_p$ approaches to a steady value at high temperature, the difference of $k$ is mainly a result from the difference of $k_e$. Equation (4) demonstrates that $k_e$ is related to the Fermi-Dirac distribution function $f_e$ and electronic transmission coefficient $T_e$. The comparison of $T_e$ and $f_e$ for $MoS_2/WS_2$ and $MoS_2/MoSe_2$ is shown in Figure 5. At $T$ = 300K, the spectra of $f_e$ demonstrates that only the electrons around the valence band edge contribute to $k_e$, while the spectra of $T_e$ demonstrates that these electrons in $MoS_2/WS_2$ and $MoS_2/MoSe_2$ have a similar transmission ability. Therefore, the $k_e$ of the two structures are nearly the same. At high temperature, however, more electrons join in the transport process. Because the $T_e$ of $MoS_2/MoSe_2$ is smaller than that of $MoS_2/WS_2$, the former has smaller $k_e$. The smaller $T_e$ of $MoS_2/MoSe_2$ is related to its

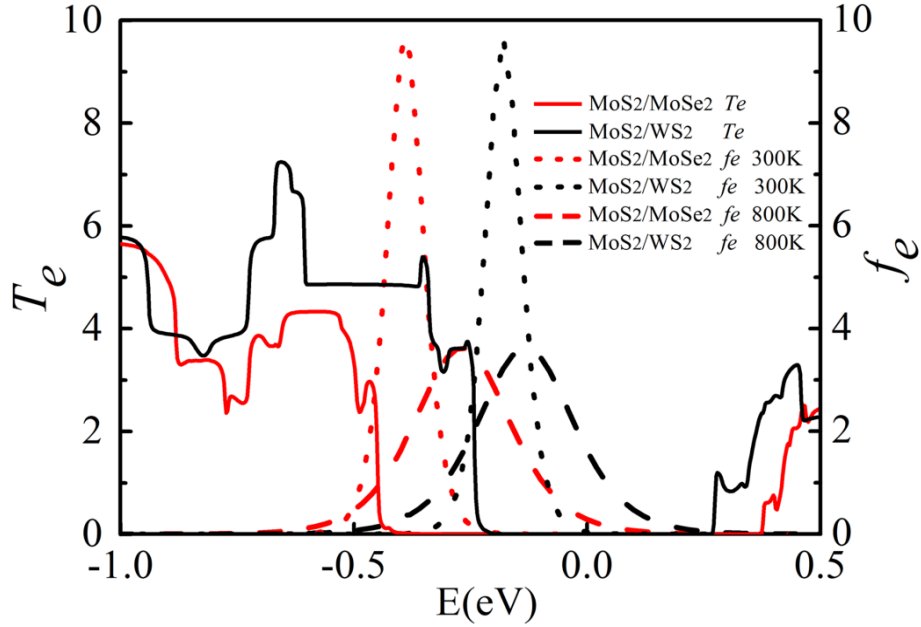

**Figure 5|** Electronic transmission coefficient $T_e$ and Fermi-Dirac distribution function $f_e$ for $MoS_2/MoSe_2$ and $MoS_2/WS_2$ with $N = 2$ as a function of electron energy $E$. The values of $\mu$ correspond to those at Max $ZT$.

……………………………………………………………………………………...

more defective interfaces originated from larger geometrical differences between $MoS_2$ and $MoSe_2$.

The above results indicate that the interfaces can enhance thermoelectric properties. Therefore, the $ZT$ of hybrid structures can be further optimized by introducing more interfaces, which can be realized by increasing the periodic number $N$ in the central scattering region while the periodic length $l$ is fixed. Figure 6 shows the Max $ZT$ values for the three hybrid nanoribbons as a function of $N$ (at $T = 800$ K). One can find that the $ZT$ values increase with $N$ and then converge to maximum values. The highest $ZT$ for $MoS_2/MoSe_2$ is 7.4, which is 2~3 times that of pristine nanoribbons.

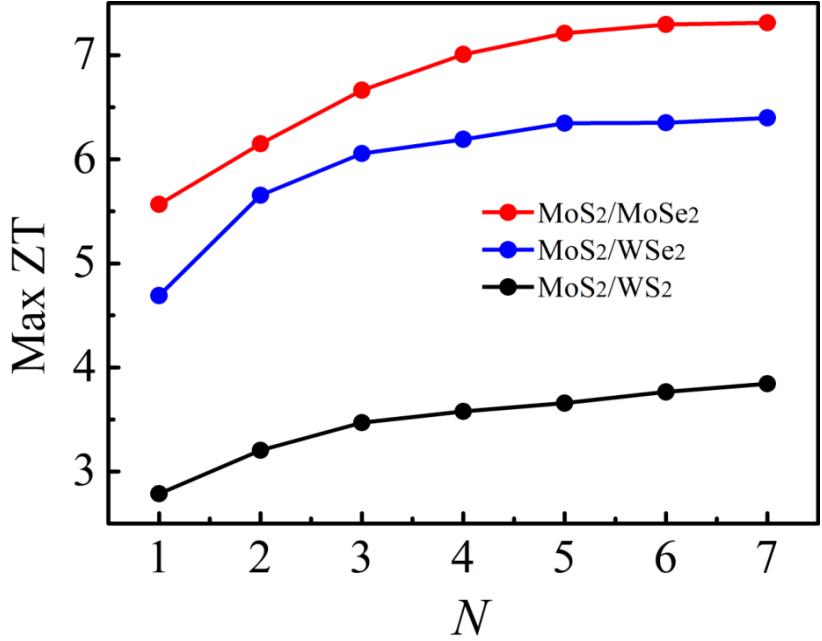

**Figure 6|** The Max *ZT* values of hybrid nanoribbons $MoS_2/MoSe_2$, $MoS_2/WSe_2$, and $MoS_2/WS_2$ with different periodic number *N*, at *T* = 800K.

............................................................................................................

To the authors' best knowledge, this *ZT* value of 7.4 is a new record of thermoelectric conversion efficiency. Table S1 in the Supplementary material shows the highest *ZT* values in different thermoelectric materials reported in the past five years. Most of the values are between 2.0 ~ 3.5. The higher *ZT* values are 5.4 and 5.5 found in black phosphorus[47] and zigzag-edged hybrid TMDs[48] respectively, which are also smaller than this discovery. It is worth noting that the *ZT* values of pristine TMDs we obtained here are lower than those reported by others. For example, the highest *ZT* value of armchair-edged $MoS_2$ nanoribbons calculated by first principles and Boltzmann transport equation is 3.5,[20] the highest *ZT* value of fewer layer $MoS_2$ calculated by first principles is 2.4.[21] The difference originates from the NEGF method because it overestimates phonon thermal conductance $k_p$. The electron mean free paths (MFP) in TMDs are 12~15 nm,[3,49] while the phonon MFPs are relatively short, about 5.3~18.1 nm.[12] The sizes of TMDs we considered vary in the range of 5 ~

15 nm. Therefore, electron transport in these nanoribbons is completely ballistic, while the phonon transport is between ballistic and diffusive regions. The NEGF method is based on the ballistic mechanism, which omits phonon scattering and thus obtains a large $k_p$. In this sense, the real *ZT* values of the pristine TMDs and their hybrid structures should be higher than what we obtained here. In addition, other functional methods including doping, edge defects, and adsorption are also helpful to further improve the *ZT* values of hybrid TMDs.[50-55] therefore, it is promising to achieve even higher *ZT* in the hybrid TMDs materials.

## IV. CONCLUSIONS

We have investigated the thermoelectric properties of hybrid armchair-edged TMD nanoribbons using the NEGF method combined with first-principles calculations and molecular dynamics simulations. The hybrid armchair nanoribbons show high *ZT* due to the fact that the hybrid interfaces reduce thermal conductance drastically. In the hybrid structures that have only three interfaces, the highest *ZT* approaches 6.1 at 800 K, while the *ZT* can be further optimized to 7.4 by more interfaces, which is a new record of high *ZT* among all the thermoelectric materials. We find that it is more efficient to improve thermoelectric properties of the hybrid interfaces by substituting X atoms than by substituting M atoms at high temperature, because the latter can pass through more electrons and thus results in a larger $k_e$. This study could be useful to design high efficient thermoelectric devices. In addition, the effect of the hybrid structures' widths on the *ZT* is very week. Although thermal conductance increases with the widths, electronic conductance also increases simultaneously, because more transport channels for electrons and phonons are opened together.

## SUPPLEMENTARY MATERIAL

See supplementary material for the SW parameters of TMDs and the highest *ZT* values in different thermoelectric materials reported in the past five years.

## ACKNOWLEDGEMENTS

This work was supported by the National Natural Science Foundation of China (Nos. 51376005 and 11474243).